\documentclass[a4paper]{llncs}

\title{Simple and Effective Relation-Based Approaches
  To XPath and XSLT Type Checking}
\subtitle{(Technical Report, Bad Honnef 2015)}
\author{Baltasar Tranc{\'o}n y
  Widemann \and Markus Lepper}
\institute{{\tt <semantics/>} GmbH}

\usepackage[bookmarks=false,colorlinks,linkcolor=blue,citecolor=blue,urlcolor=blue]{hyperref}
\usepackage{pgf}
\usepackage{amsmath,amssymb}
\usepackage{xcolor}
\usepackage{fancyhdr}
\usepackage{hyperref}

\newcommand\xpath[1]{\texttt{#1}}
\newcommand\xml[1]{\texttt{#1}}
\newcommand\inv{\breve{\enspace}}
\newcommand\Types{\ensuremath{\mathbb T}}
\newcommand\Elems{\ensuremath{\mathbb E}}
\newcommand\Attrs{\ensuremath{\mathbb A}}
\newcommand\Power{\ensuremath{\mathcal P}}
\newcommand\Content{\ensuremath{\mathcal C}}
\newcommand\rel[1]{\ensuremath{\mathsf{#1}}}
\newcommand\dom{\ensuremath{\mathrm{dom}}}

\renewcommand\emptyset{\varnothing}
\newcommand\infrule[1]{\ensuremath{(\mathsf{R}#1)}}

\begin{document}

\maketitle

\begin{abstract}
  XPath is a language for addressing parts of an XML document. We give
  an abstract interpretation of XPath expressions in terms of
  relations on document node types. 
  Node-set-related XPath
  language constructs are mapped straightforwardly onto basic,
  well-understood and easily computable relational operations. Hence
  our interpretation gives both extremely concise type-level
  denotational semantics and a practical analysis tool for the
  node-set fragment of the XPath\,1.0 language. This method is part of 
  the TPath implementation of XPath.

  XSL-T is a pure functional language for transforming XML documents.
  For the most common case, the transformation into an XML document,
  type checking of the transformation code is unfeasible in general, but strongly
  required in practice. It turned out that the relational approach of TPath
  can be carried over to check all fragments of the result language, which are
  contained verbartim in the transformation code. This leads to a technique
  called ``Fragmented Validation'' and is part of the txsl implementation of XSL-T.

\end{abstract}

\section{Introduction}


In the context of the authors' {\sffamily metatools} compiler construction
toolkit, XML\cite{xml11} plays a fundamental role for the encoding of 
data structures as well as for text documents. 
Related standardized  languages are XPath\cite{xpath10} 
for inquiring XML objects, and XSL-T\cite{xslt} transformation.
In this context, constraints on the valid syntactic structure of classes of documents are 
defined by ``Document Type Definitions'', DTD\cite{xml11}.
It turned out that all these components had to be implemented from scratch,
for modular and compositional usage, and satisfying error diagnosis.

Given a fixed DTD and a fixed XPath expression, the two 
most important questions in practice are 
\emph{satisfiability} of the expression, when applied to node of a certain kind, 
and the \emph{general type} of the resulting nodes;
given an XSL-T program and two DTDs, 
the question is for the \emph{correctness} of the program, i.e.\ will the result
always adhere to the second DTD if the input adheres to the first.

On the theoretical side, much attention has been given to these and
related questions as \emph{exact} problems: in general they turn out
to be undecidable.  A hierarchy of syntactic \emph{restrictions} of
the two languages involved gives decidable subsets with complexity
ranging from P to NEXPTIME; see the exhaustive discussions in
\cite{benedikt08,Geneves:2007:ESA:1273442.1250773}.

Here we propose the opposite approach, namely to give
\emph{approximate} solutions, in the tradition of type systems, for
the \emph{full} expressiveness of DTD, XPath~1.0 and XSL-T~1.0.  
Semantically
problematic constructs are supported with trivial approximations for
graceful degradation of the analysis, rather than rejected
categorically.  
Thus we give a light-weight pragmatic solution which answers the
practically most relevant, structural questions.

\section{Abstract Interpretation of DTD and XPath}

Our proposed analysis of a particular combination of DTD and XPath 
has the form of a simple, easily understood and just as easily implemented
algorithm, which calculates an 
\emph{upper bound} of possible results by
\emph{abstract interpretation} based on
a \emph{relation algebra}.
The formal presentation given here  corresponds
almost literally to our concrete implementation. It uses a
\textsf{Java} library of declarative finite set-based relational
operations that is part of our {\sffamily metatools} toolkit.

The chief purpose and obvious interpretation of an XPath expression is
to select a subset of the nodes of a \emph{concrete} XML document. At a higher level
of abstraction, the same XPath expression can be interpreted as
selecting a subset of the nodes of \emph{all} XML documents conforming to a
fixed DTD. By partitioning such
an infinite set of potential nodes into the finite set of \emph{node types} in the sense of \cite[Sect.\ 5]{xpath10}, an
abstract interpretation of the XPath language can be given that
assigns an upper bound of selectable node types to each XPath
expression.  This interpretation relates the types of
context nodes to the types of nodes potentially selected by the
expression from that context.  

All of the above general questions can be addressed thus in a uniform
way; for instance, the empty set of selected node types implies
unsatisfiability of the interpreted expression.  XML toolkits can
leverage the node type information for the development, analysis,
optimization, maintenance and quality assurance of XML processing
applications.

Abstract interpretation of the XPath language is far simpler than for
the average programming language because of the absence of recursive
expressions.  No fixpoint computations (cf.~\cite{cousot:absint}) are
needed; our interpretation is purely syntax-directed and bottom-up.

\subsection{Node Types and XPath Axes}
\label{axes}

For one fixed DTD we introduce a finite
set $\Types$ of node types, which partition the nodes of all documents
conforming to that DTD. For each node type, there is a primitive,
characteristic XPath expression that selects nodes of this type, but
excludes all others.  All relations under consideration are binary
relations on $\Types$.  Table~\ref{tab:generic} shows the four
generic node types that apply to all DTDs.

\begin{table}[t]
  \caption{Generic node types}
  \label{tab:generic}
  \centering
  \begin{tabular}{c@{\quad}c@{\quad}c}
    \hline
    \textbf{XPath Node Type} & \textbf{Characteristic Expression} & \textbf{Type Symbol}
    \\ \hline
    \xpath{root} & \xpath{/} & $\rho$
    \\
    \xpath{comment} & \xpath{comment()} & $\gamma$
    \\
    \xpath{processing instruction} & \xpath{processing-instruction()} & $\pi$
    \\
    \xpath{text} & \xpath{text()} & $\tau$
    \\ \hline
  \end{tabular}
\end{table}

Besides the four generic node types, each DTD declares a finite number
of element and attribute names, each giving rise to a node type.  The
characteristic XPath expressions for an element named $e$ and an
attribute named $a$ are \xpath{child::$e$} and \xpath{attribute::$a$}.
These can be abbreviated to \xpath{$e$} and \xpath{@$a$},
respectively, which we shall also use to denote the corresponding node
type.  We write $\Elems$ and $\Attrs$ for the corresponding sets of
types, hence $\Types = \{ \rho, \gamma, \pi, \tau \} \cup \Elems \cup
\Attrs$.

In XPath, navigation within a document is accomplished by composing
steps classified by so-called \emph{axes}.  The specification text
defines \emph{thirteen} axes, which can be reduced to a basis
of \emph{three} (see Table~\ref{tab:axes}): $\rel c$ and $\rel f$
model the nesting and local order of element nodes, respectively,
whereas $\rel @$ models the placement of attributes.  All three primitive axes are specific to a fixed DTD.  The following paragraphs consider them in turn.

\begin{table}[t]
  \caption{XPath axes}
  \label{tab:axes}
  \centering
  \begin{tabular}{c@{}c@{\enspace}c@{\quad}c@{}c@{\enspace}c}
    \hline
    \textbf{XPath Axis} & \multicolumn{2}{c}{\textbf{Relation}} &
    \textbf{XPath Axis} & \multicolumn{2}{c}{\textbf{Relation}}
    \\
    \textbf{Name} & \textbf{Name} & \textbf{Definition} &
    \textbf{Name} & \textbf{Name} & \textbf{Definition}
    \\ \hline
    \xpath{child} & $\rel c$ & --- &
    \xpath{parent} & $\rel p$ & $(\rel c \cup \rel @)\inv$
    \\
    \xpath{attribute} & $\rel @$ & --- &
    \xpath{self} & $\rel s$ & $I_\Types$
    \\
    \xpath{following-sibling} & $\rel f$ & --- &
    \xpath{preceding-sibling} & $\rel r$ & $\rel f\inv$
    \\
    \xpath{descendant} & $\rel d$ & $\rel c^+$ &
    \xpath{ancestor} & $\rel a$ & $\rel p^+$
    \\
    \xpath{descendant-or-self} & $\rel D$ & $\rel c^*$ &
    \xpath{ancestor-or-self} & $\rel A$ & $\rel p^*$
    \\
    \xpath{following} & $\rel F$ & $\rel A \circ \rel f\circ \rel D$ &
    \xpath{preceding} & $\rel R$ & $\rel A \circ \rel r \circ \rel D$
    \\
    \xpath{namespace} & \multicolumn{2}{c}{(deprecated)}
    \\ \hline
  \end{tabular}
\end{table}

\subsubsection{Child Axis}

The relation $\rel c$ is the smallest relation that
meets the following requirements:
\begin{enumerate}
\item The root type is related to the types of all elements admissible
  as document (outermost) elements, and to the comment and processing
  instruction node type.  Since the DTD formalism cannot express
  constraints on the admissible elements, generally $\{ \rho \} \times
  \bigl(\Elems \cup \{ \gamma, \pi \}\bigr) \subseteq \rel c$.

  Informal normative constraints (such as XHTML\,1.0 allowing only
  \xml{html} as the root element) can be imposed to improve the
  analysis.
\item Any element type is related to the comment and processing
  instruction node types, so $\Elems \times \{ \gamma, \pi \}
  \subseteq \rel c$.
\item The type of any element declared by the DTD with mixed content
  is related to the text node type.  Let $\Elems_{\text m}$ be the types of all
  elements declared in the form \xml{<!ELEMENT $e$ (\#PCDATA |
    $\dots$)*>}. Then $\Elems_{\text m} \times \{ \tau \} \subseteq \rel c$.
\item The type of any element declared by the DTD is related to the
  types of elements occurring in its content declaration, whether of
  mixed or element content (regular) type. Note that the content
  specifier \xml{ANY} is not supported, since it defies closed-world
  static analysis by definition.
\end{enumerate}

\subsubsection{Attribute Axis}

The relation $\rel @$ is the smallest relation that
relates the type assigned to each element declared by the DTD to the
types assigned to the attributes declared for that element by the DTD.
That is, for every pair of matching declarations
\begin{align*}
  \xml{<!ELEMENT $e$ $\dots$>} && \xml{<!ATTLIST $e$ $\dots$ $a$ $t$
    $v$ $\dots$>}
\end{align*}
that declare an attribute $a$ of element $e$ with value type $t$ and
default value $v$, there is a pair $(e, @a) \in \rel @$.

\subsubsection{Following-Sibling Axis}

The \xpath{following-sibling} axis is by far the most complicated of
the primitive XPath axes. The following paragraphs construct a
relational interpretation in four conceptual steps:
\begin{enumerate}
\item Define an even more primitive relation $\rel n_e$
  (\emph{next-proper-sibling}) per element $e$ by induction on the
  declared content model.
\item Define a relation $\rel f_e$ per element $e$ that includes both
  the transitive closure of $\rel n_e$ and potentially intervening
  non-element node types. 
\item Demonstrate that it is infeasible to handle siblings on a
  per-element basis.
\item Conclude by taking the union of all local relations $\rel f =
  \bigcup_{e \in E} \rel f_e$ as a reasonable approximate
  interpretation.
\end{enumerate}

\subsubsection{Local Next-Proper-Sibling Relation}
\label{next-proper-sibling}

We construct a relation $\rel n_e$ per element $e$ with the following
meaning: $(t_1, t_2) \in \rel n_e$ iff an element-or-text node of type
$t_1$ may be followed immediately by an element-or-text node of type
$t_2$ within the content of an element of type $e$.

For mixed content, this construction is trivial: For an element
declared in the form \xml{<!ELEMENT $e$ (\#PCDATA | $e_1$ | $\dots$ |
  $e_n$)*>}, take the set of node types $M_e = \{ \tau, e_1, \dots,
e_n\}$ and let $\rel n_e = (M_e \times M_e) \setminus I_{\{\tau\}}$,
since the only combination of consecutive children forbidden by the
XPath data model is a pair of text nodes.

For so-called element content, the construction is more complicated.
Consider the following mathematical interpretation of DTD content
models, as implied in the XML specification: Each content model $c \in
\Content$ denotes a \emph{content language} $[c] \subseteq \Elems^*$
(set of finite sequences of element types), by the usual
interpretation of regular expressions; see Table~\ref{tab:regexp}.
For compositionality we add the empty specification \xml{()}, which had
been forgotten in the XML specification.  As
usual, $\varepsilon$ denotes the empty word.  Confer also DTD normalization in \cite{benedikt08}.

\begin{table}[t]
  \caption{Interpretation of DTD content models as regular languages}
  \label{tab:regexp}
  \vspace{-\baselineskip}
  \begin{align*}
    [\xml{($c_1$, $\dots$, $c_n$)}] &= [c_1] \cdot \ldots \cdot [c_n] &
    [\xml{$c$?}] &= [c] \cup [\xml{()}] &
    [e] &= \{ e \}
    \\
    [\xml{($c_1$ | $\dots$ | $c_n$)}] &= [c_1] \cup \dots \cup [c_n] &
    [\xml{$c$*}] &= [\xml{$c$+}] \cup  [\xml{()}] &
    [\xml{()}] &= \{\varepsilon\}
    \\
    A \cdot B &= \{ \alpha\beta \mid \alpha \in A; \beta \in B \} &
    [\xml{$c$+}] &= \bigcup_{n=1}^\infty [c]^n &
    A^n &= \underbrace{A \cdot \ldots \cdot A}_n
  \end{align*}
\end{table}
  
We say that a content language $A \in \Types^*$ \emph{ends with} $P \subseteq
\Types$ iff $P$ is the smallest set of node types such that all
sequences $\alpha \in A \setminus \{\varepsilon\}$ end with some $t
\in P$.

We calculate the next-sibling relation inductively over the
meta-language $\Content$.  To this end, we
shall define an auxiliary function $m : \Content \times \Power(\Types)
\to \Power(\Types \times \Types) \times \Power(\Types)$, such that for
$(c, P) \stackrel{m}{\mapsto} (S, Q)$ the following invariants hold:

\begin{enumerate}
\item The relation $S$ contains precisely the pairs $(t, u)$ such
  that, for all content languages $A$ ending with $P$, $tu$ occurs as
  a contiguous subsequence in $A \cdot [c]$.  (Since the language of
  singletons from $P$ is a language ending with $P$, at least $u$ must
  stem from $[c]$.)
\item For all content languages $A$ ending with $P$, $A \cdot [c]$
  ends with $Q$.
\end{enumerate}
Note that the two invariants together fix the result of $m$, which
depends on $c$ only via $[c]$: We define two equivalence relations as
$c \cong c'$ iff $[c] = [c']$ (\emph{modulo language}), and $c \approx
c'$ if and only if $m(c, P) = m(c', P)$ for all $P$ (\emph{modulo
  abstract interpretation}), respectively.  Then we have the lemma
$(\cong) \subseteq (\approx)$, which will be used below.

Intuitively $m$ models the operation of a DTD content model on a set
of prefixes, namely updating the set of possible final elements, and
extending the set of possible neighbourships.  For an element declared
as \xml{<!ELEMENT $e$ $c$>}, we then define $\rel n_e = S$ iff $(c,
\emptyset) \stackrel{m}{\mapsto} (S, Q)$ for some $Q$.

\begin{table}[t]
  \caption{Induction rules for next proper siblings in element content}
  \label{tab:match}
  \vspace{-\baselineskip}
  \begin{align*}
    \begin{gathered}
      (e, P) \stackrel{m}{\mapsto} (P \times \{e\}, \{e\}) \quad \infrule1
      \\[1ex]
      (\xml{()}, P) \stackrel{m}{\mapsto} (\emptyset, P) \quad \infrule2
      \\[1ex]
      \frac{\bigl(\xml{($c$, $c$?)}, P\bigr) \stackrel{m}{\mapsto} (S,
        Q)}{(\xml{$c$+}, P) \stackrel{m}{\mapsto} (S, Q)} \quad \infrule3
    \end{gathered}
    &&
    \begin{gathered}
      \frac{(c_1, P) \stackrel{m}{\mapsto} (S, Q) \quad (c_2, Q)
        \stackrel{m}{\mapsto} (T, R)} {(\xml{($c_1$, $c_2$)}, P)
        \stackrel{m}{\mapsto} (S \cup T, R)} \quad \infrule4
      \\[1ex]
      \frac{(c_1, P) \stackrel{m}{\mapsto} (S, Q) \quad (c_2, P)
        \stackrel{m}{\mapsto} (T, R)} {(\xml{($c_1$ | $c_2$)}, P)
        \stackrel{m}{\mapsto} (S \cup T, Q \cup R)} \quad \infrule5
    \end{gathered}
  \end{align*}
\end{table}

Table~\ref{tab:match} shows the induction rules that define $m$ and
induce a syntax-directed, deterministic algorithm.  To verify
that these rules are adequate and complete, we systematically check the
invariants given above for all syntactic cases:
\begin{enumerate}
\item The single-element case $e$ \infrule1 is trivial.
\item The nullary sequence case \xml{()} \infrule2 is equally trivial.
  Together with the choice rule \infrule5 it defines the rules for the
  content model iterators \xml{?}  and \xml{*}.
\item The binary cases \xml{($c_1$, $c_2$)} \infrule4 and \xml{($c_1$ |
    $c_2$)} \infrule5 formalize the notion of sequence and choice,
  respectively.  Ternary and higher cases follow uniquely because of
  associativity modulo language.
\item The most interesting case is the iterator \xml{+} \infrule3.  It
  holds asymptotically that
  \begin{equation*}
    \xml{$c$+} \cong \xml{($c$ | ($c$, $c$) | ($c$, $c$, $c$) | $\dots$)}
  \end{equation*}
  It is easy to show inductively that $\xml{($c$, $c$)} \approx
  \xml{($c$, $c$, $c$)}$ and hence, by associativity, all sequences of
  two or more $c$s are equivalent. We conclude
  \begin{equation*}
    \xml{$c$+} \approx \xml{($c$ | ($c$, $c$))} \cong \xml{($c$, $c$?)}
  \end{equation*}
  which justifies the rule.

  Note that the given rule invokes the body $c$ twice, and that these
  formulas directly represent the operation of our algorithm.  So the
  there may be exponential worst-case complexity for nested \xml{+}
  and \xml{*} iterators. To our experience, practical content models
  are not nested deeply enough to cause trouble in this regard.  The
  XHTML~DTD, for instance, does not contain any irreducible nested
  iterators at all.
\end{enumerate}

\subsubsection{Local Following-Sibling Relation}

The following-sibling relation can be derived from the
next-proper-sibling relation by noting that additionally, transitivity
is required, and comments and processing instructions may intervene
arbitrarily.  Hence we define $\rel f_e = \rel n_e^+ \cup (\Elems_e
\times Z)^\diamond \cup (Z \times Z)$, where $\Elems_e$ is the set of
element node types occuring in $\rel n_e$ and $Z = \{ \gamma, \pi
\}$. The analog case for the children of the root node is $\rel f_\rho
= (\Elems \times Z)^\diamond \cup (Z \times Z)$.  The operator
$^\diamond$ denotes symmetric closure of a relation: $A^\diamond = A
\cup A\inv$.

\subsubsection{Per-Element Following Siblings}

Consider for a moment the impact of handling the following-sibling
relation $\rel f_e$ seperately for each element type $e$: The type $t$
of a node is not enough to infer its following siblings. The type $e$
of its parent is required as well and, because of the ancestor axis,
so are the types of all of its ancestors. Hence we are forced to lift
our interpretation from context-free relations $r \subseteq \Types
\times \Types$ on node types to context-sensitive relations $\overline
r \subseteq \Types^* \times \Types^*$ on node type \emph{paths}.  On
the upside, this can be done in a mathematically straightforward way.
For every context-free primitive relation there is a corresponding
context-sensitive one
\begin{align*}
  \frac{(t, u) \in \rel c}{(\alpha t, \alpha tu) \in \overline{\rel c}} &&
  \frac{(t, u) \in \rel @}{(\alpha t, \alpha tu) \in \overline{\rel @}} &&
  \frac{(t, u) \in \rel f_e}{(\alpha et, \alpha eu) \in \overline{\rel f_e}}
  \end{align*}
  where $\alpha \in \Types^*$ is any (possibly empty) sequence of
  ancestor node types. On the downside, this lifting ruins the simple
  finite representation of relations: since $\alpha$ can take on
  infinitely many values, the structure of composite relations becomes
  quite complicated; for instance consider
\begin{equation*}
  \overline{\rel F} = \bigl((\overline{\rel c} \cup
  \overline{\rel @})\inv\bigr)^* \circ \bigcup_{e \in E}
  \overline{\rel f_e} \circ (\overline{\rel c} \cup \overline{\rel @})^*
\end{equation*}
Though finding an effective representation for this kind of relations
may be an interesting challenge, we leave it as an open problem for
now.

\subsubsection{Unified Following-Sibling Relation}

Having conceded that there is no obvious solution to the context
problem for the following-sibling relations, we simply define
\begin{equation*}
  \rel f = \rel f_\rho \cup \bigcup_{e \in E} \rel f_e
\end{equation*}
thereby abstracting from the possibility of an element having
significantly different potential siblings in different contexts. It
remains to be established empirically how much information is lost in
this way.  Note that XPath subexpressions which do not use the
horizontal axes are not affected.

\subsection{Abstract Interpretation of XPath Expressions}

Table~\ref{tab:absint} shows the abstract interpretation, which is a
partial function from XPath expressions to node type relations,
specified as a relation $\leadsto$. The details are explained in the
following subsections.

\begin{table}[t]
  \caption{Syntax-directed abstract interpretation rules (summary)}
  \label{tab:absint}
  \vspace{-\baselineskip}
  \begin{gather*}
    \frac{a \leadsto A \quad t \leadsto T}{\xpath{/$a$::$t$} \leadsto (\Types
      \times \{ \rho \}) \circ A \circ T} \qquad \frac{a \leadsto A \quad t
      \leadsto T}{\xpath{$a$::$t$} \leadsto A \circ T} \qquad \frac{e \leadsto
      E \quad a \leadsto A \quad t \leadsto T}{\xpath{$e$/$a$::$t$} \leadsto E
      \circ A \circ T} \\[1ex]
    \begin{aligned}
      \frac{e \leadsto E \quad f \leadsto F}{\xpath{$e$|$f$} \leadsto
        E \cup F} &\quad& \frac{e \leadsto E \quad f \leadsto
        F}{\xpath{$e$[$f$]} \leadsto E \circ I_{\dom(F)}} &\quad&
      \frac{e \leadsto E \quad f \not\in \dom(\leadsto)}{\xpath{$e$[$f$]} \leadsto
        E}
    \end{aligned}
  \end{gather*}
\end{table}

\subsection{Location Paths}

An XPath location path expression takes one of three forms: absolute
(\xpath{/$a$::$t$}), relative (\xpath{$a$::$t$}) or recursive
(\xpath{$e$/$a$::$t$}); following filter predicates will be considered
below.  Interpretation is defined as follows:
\begin{enumerate}
\item Compute a base relation $E$: In the absolute case, the context
  is ignored and replaced by the document root; set $E = \Types \times
  \{\rho\}$. In the relative case, set $E = I_\Types$.  In the
  recursive case, recursively compute the relation $E$ assigned to
  $e$. If undefined, the interpretation of the whole expression is
  undefined.
\item Otherwise, assign a relation $A$ to the axis $a$ (see
  Sect.~\ref{axes}) and a relation $T$ to the test $t$ (see below).
\item Assign the relation $E \circ A \circ T$ to the whole expression.
  In the relative case, this simplifies to $A \circ T$.
\end{enumerate}

\noindent Table~\ref{tab:tests} shows the relations assigned to
generic node tests. Name tests are mapped to relations as follows:
\begin{enumerate}
\item An explicit name test $n$ maps to the relation $I_{\{n\}}$ if
  the principal node type for the current axis $a$ is \xpath{element},
  or to $I_{\{@n\}}$ if the principal node type is \xpath{attribute}.
  The latter applies to the attribute axis only, the former to all
  other axes except the unsupported namespace axis.
\item A wildcard name test \xpath{*} maps to the relation $I_\Elems$
  or $I_\Attrs$, if the principal node type for the current axis $a$
  is \xpath{element} or \xpath{attribute}, respectively.
\end{enumerate}

\begin{table}[t]
  \caption{Generic node tests}
  \label{tab:tests}
  \centering
  \begin{tabular}{c@{\quad}c@{\qquad}c@{\quad}c}
    \hline
    \textbf{XPath Node Test} & \textbf{Relation} &
    \textbf{XPath Node Test} & \textbf{Relation}
    \\ \hline
    \xpath{node()} & $I_\Types$
    \\
    \xpath{text()} & $I_{\{\tau\}}$ &
    \xpath{processing-instruction()} & $I_{\{\pi\}}$
    \\
    \xpath{comment()} & $I_{\{\gamma\}}$ &
    \xpath{processing-instruction($\dots$)} & $I_{\{\pi\}}$
    \\ \hline
  \end{tabular}
\end{table}

\subsection{Unions}

An XPath union expression has the form \xpath{$e$|$f$}. 
Interpretation is defined as follows:
\begin{enumerate}
\item Assign a relation $E$ to the left argument $e$.  Likewise,
  assign a relation $F$ to the right argument $f$.  If either is
  undefined, the interpretation of the whole expression is undefined.
\item Otherwise, assign the relation $E \cup F$ to the whole
  expression.
\end{enumerate}

\subsection{Filters}

An XPath filter expression has the form \xpath{$e$[$f$]}.
Interpretation is defined as follows:
\begin{enumerate}
\item Assign a relation $E$ to the base expression $e$. If undefined,
  the interpretation of the whole expression is undefined.
\item Otherwise, assign a relation $F$ to the filter predicate $f$
  (see below).  If undefined, assign the relation $E$ to the whole
  expression. This cop-out interpretation is safe as an upper bound
  because a filter can at most remove node types from the base node
  set.
\item Otherwise, assign the relation $E \circ I_{\dom(F)}$ to the
  whole expression. This interpretation is safe as an upper bound
  because a filter predicate that evaluates to a node set $s$ in the
  context of a node $x$ by definition selects $x$ iff $s$
  is nonempty. This in turn implies that the type $t$ of $x$ is
  related by $F$ to the types of the members of $s$, hence $t \in
  \dom(F)$.
\end{enumerate}

\noindent Filter predicates are mapped to relations as follows:
\begin{enumerate}
\item The default is the ordinary abstract interpretation of $f$, if
  defined.
\item Logical operators are treated specially:
  \begin{enumerate}
  \item A predicate of the form \xpath{$f$ and $g$}, where the relations
    $F$ and $G$ are assigned recursively to $f$ and $g$, respectively,
    is mapped to the relation $F \cap G$.
  \item A predicate of the form \xpath{$f$ or $g$}, where the relations
    $F$ and $G$ are assigned recursively to $f$ and $g$, respectively,
    is mapped to the relation $F \cup G$.
  \end{enumerate}
  Note that this treatment is not strictly necessary, because if both
  $f$ and $g$ evaluate to node sets in the context selected by $e$,
  then the expressions \xpath{$e$[$f$ and $g$]} and \xpath{$e$[$f$ or
    $g$]} are equivalent to \xpath{$e$[$f$][$g$]} and
  \xpath{$e$[$f$|$g$]}, respectively.
\end{enumerate}
For all other filter predicates, the abstract interpretation is
undefined, and hence does not impose any restriction on the relation
assigned to the base expression.  Note that filter predicates of the
form \xpath{not($f$)} are explicitly not covered, as the
anti-monotonicity of negation would break the upper bound property of
the abstract interpretation.

\subsection{Other Expressions}

The XPath function $\xpath{id}$ selects elements from the whole
context document, regardless of the particular context node, by the
value of their identity attribute. A function call expression of the
form \xpath{id($x$)} is mapped to the relation $\Types \times
\Elems_{\text i}$, where $\Elems_{\text i}$ is the set of all element
node types in the DTD that have a declared attribute of value type
\xml{ID}.

Otherwise, our abstract interpretation is undefined for all XPath
expressions not covered above.  Similar restrictions apply to other
semantical models of XPath such as
\cite{Marx:2005:SCN:1083784.1083792} as well, and do not hinder
analysis of document structure unduly.

\subsection{Semantic Properties and Their Applications}
\label{theorems}

Consider some fixed XPath expression $e$ and DTD. If the abstract
interpretation assigns a relation $E$ on node types to $E$, then the
following properties should hold:
\begin{enumerate}
\item For each valid document with respect to the DTD and each context
  node within the document, $e$ either selects a node set from the
  document or fails to evaluate, but does not evaluate to a string,
  number, truth value or other data (\emph{well typing}).
\item Let $D$ be a nonempty set of node types and $R = ED$ the image
  of $D$ under the relation $E$.
  \begin{enumerate}
  \item For absolute $e$, $R$ does not depend on $D$.
  \item For both absolute and relative $e$, each node in the set
    selected by $e$ starting from a context node of some type $s \in
    D$ has a type in $t \in R$ (\emph{completeness}).
  \end{enumerate}
\end{enumerate}

As a corollary of completeness, if $R$ is empty then $e$ is
unsatisfiable: the expression selects only the empty node set from any
valid document and any context node of type $s \in D$. This has
important practical implications:
\begin{enumerate}
\item For the root of an XPath expression, it is most likely an error
  and should be reported to the user of the XML processing tool.
\item For any expression fragment, it indicates optimization potential
  in the XPath implementation: The most frequently-used axes are the
  \xpath{child} and \xpath{attribute} axes, as witnessed by their
  special abbreviated syntax. Node selection along these axes is
  usually implemented by recursive tree traversal, for instance using
  the \emph{visitor} style pattern. The current state of the traversal
  is specified by a relative XPath subexpression. Whenever the type of
  the root node of a subtree is not in the image of the relation
  associated with the governing subexpression, traversal of the whole
  subtree can be pruned safely.
\end{enumerate}

Note that there is no \emph{soundness} property dual to completeness:
One might expect that for each type $t \in R$, there is a valid
document and context node such that $e$ selects a node of type $t$.
But since our interpretation is an approximate upper bound, this is
not the case in general.

\subsection{Usage of the Command Line Tool}

The following inputs to the command line tool show typical questions to
a particular DTD, here XHTML~1.0\cite{xhtml10}:

\begin{itemize}
\item
{\tt make test XPATH="p/ol"}\\
{\footnotesize $\Longrightarrow$ {\tt child::p/child::ol}}\\
$\Longrightarrow$ \verb!{}!\\
(``Can a \texttt{p} element ever contain directly an \texttt{ol} element?'' -- ``No!'')
\item
{\tt make test XPATH="p//ol"}\\
{\footnotesize $\Longrightarrow$ {\tt child::p/descendantOrSelf::node()/child::ol}}\\
$\Longrightarrow$ {\tt \verb!{!(form,ol),(ins,ol),(map,ol),(body,ol),(fieldset,ol), \\
(object,ol),(li,ol),(del,ol),(dd,ol),(td,ol),(blockquote,ol),\\
(th,ol),(button,ol),(div,ol),(noscript,ol)\verb!}!}\\
(``Can a \texttt{p} element contain indirectly an \texttt{ol} element?'' -- ``Yes!
And the \texttt{p} element itself is contained in a \texttt{form} or an \texttt{ins} or a 
\texttt{map} \ldots '')
\item
{\tt make test XPATH="self::p//*[ol]"}\\
{\footnotesize$\Longrightarrow$\tt {self::p/descendantOrSelf::node()/child::*[child::ol]}}\\
$\Longrightarrow$ {\tt\verb!{!(p,fieldset),(p,del),(p,td),(p,ins),(p,li),(p,button),\\
(p,noscript),(p,dd),(p,th),\ldots\verb!}!}\\
(``Which element under that \texttt{p} element can directly contain that \texttt{ol}?'')
\end{itemize}

\section{XSL-T and Fragmented Validation}

When trying to apply the standard open source XSL-T implementation ``Xalan'' \cite{xalan},
it soon turned out that error diagnosis is too bad for efficient programming work. So
we decided to implement our own XSLT~1.0 processor. It is based on the
``tdom'' Typed Document Model, which generates a collection of Java classes from a
DTD, for serialization, deserialization, construction and inquiry in  
a \emph{strictly typed} fashion.\cite{tlw03}

In this setting, an XSL-T program is a collection of trees of two different 
``colors'', namely the pure XSL-T code, and the sub-trees from the result language,
which are interspersed in the code, and which will be combined later, when the
code is applied to some input, to construct the output document.
In our implementation, the connection between the leaves of a tree of one color
and the root of the tree of the other color are realized non-invasively, by an
adjoined map, because tdom, being strictly typed, can per se not express connections of this
mixed nature. Then simply a visitor class must be constructed which respects this
map and calls the other two visitors (generated by tdom) accordingly, to gain all 
the comfort of tdom declarative programming.

It soon turned out that already when parsing the transformation source, i.e.\ when
constructing the internal model of the code, the target fragments can easily be
validated against the target DTD. Two kinds of non-determinism come into play:

First: Whenever a fragment starts with a reference to a target DTD element, 
\emph{all} positions in \emph{all} target content models must be considered, 
because the later context is not known.

Second: Whenever XSL-T code is interspered into a target DTD fragment, the 
transitive closure of the sibling relation must be taken for all valid transitions,
because the XSL-T code may produce zero to all of all those elements still
missing to complete the current content model.

This technique is explained in detail in \cite{rtafragval}. It is easily implemented
when the parsing process is again based on \emph{relations}.
That this technique comes from 
the preceding XPath research can be seen clearly when comparing Table~1 there with
Table~\ref{tab:match} above.
It turned out to be very efficient in comparative tests and very helpful in practice.

\begin{flushleft}
\small
\bibliographystyle{splncs}
\bibliography{xpath,../xslt/xslt.bib}
\end{flushleft}

\end{document}